\documentclass{ws-ijmpd}

\begin{document}

\markboth{P. S. Negi}
{NEUTRON STAR SEQUENCES}
\catchline{}{}{}{}{}

\title{NEUTRON STAR SEQUENCES AND THE STARQUAKE GLITCH MODEL FOR THE CRAB AND THE VELA PULSARS\\
}

\author{\footnotesize P. S. Negi
}

\address{Department of Physics, Kumaun University, Nainital - 263002, India\\
psnegi\_nainital@yahoo.com}



\maketitle
\begin{history}
\received{Day Month Year}
\revised{Day Month Year}
\comby{Managing Editor}
\end{history}


\begin{abstract}
We construct for the first time, the sequences of stable neutron star (NS) models 
capable of explaining simultaneously, the glitch healing parameters, $Q$, of both the 
pulsars, the Crab ($Q \geq 0.7$) and the Vela ($Q \leq 0.2$), on the basis of starquake
 mechanism of glitch generation, whereas the conventional NS models cannot give such consistent explanation.
 Furthermore, our models also yield an upper bound on NS masses similar to those obtained in the literature for a variety of modern equations of state (EOSs) compatible with causality and dynamical stability. If the lower limit of the observational constraint of (i) 
 $Q \geq 0.7$ for the Crab pulsar and (ii) the recent value of the moment 
 of inertia for the Crab pulsar (evaluated on the basis of time-dependent acceleration 
 model of 
 the Crab Nebula)
 , $I_{\rm Crab,45} \geq 1.93$ 
 (where $I_{45}=I/10^{45}\,{\rm g.cm}^2$), both are  imposed together on 
our models, the models yield the value of matching density, 
$E_b = 9.584 \times 10^{14}{\rm\,g\,cm}^{-3}$ at the core-envelope boundary. 
This value of matching density yields a model-independent upper bound on neutron star 
masses, $M_{\rm max} \leq 2.22 M_\odot$, and the strong lower bounds on surface 
redshift $z_R \simeq 0.6232$ and mass $M \simeq 2.11 M_\odot$ for the Crab ($Q \simeq 
0.7$) and the strong upper bound on surface redshift $z_R \simeq 0.2016 $, mass $M 
\simeq 0.982 M_\odot$ and the moment 
 of inertia  $I_{\rm Vela,45} \simeq 0.587$ for the Vela ($Q \simeq 0.2$) pulsar. However, for the 
observational constraint of the `central' weighted mean value $Q \approx
0.72$, and $I_{\rm Crab,45} > 1.93$, for the Crab pulsar, the minimum surface redshift 
and mass of the Crab pulsar are slightly increased to the values $z_R \simeq 0.655$ 
and  $M \simeq 2.149 M_\odot$ respectively, whereas corresponding to the `central' weighted mean value 
$Q \approx 0.12$ for the Vela pulsar, the maximum surface redshift, mass and the moment 
 of inertia for the Vela
 pulsar are slightly decreased to the values $z_R \simeq 0.1645,\, M \simeq 0.828 
 M_\odot$ and $I_{\rm Vela,45} \simeq 0.459$ respectively. These results set an upper and lower bound on the energy of a 
gravitationally redshifted electron-positron annihilation line in the range of about 
0.309 - 0.315 MeV from the Crab and in the range of about 0.425 - 0.439 MeV from the 
Vela pulsar.
\end{abstract}

\keywords{Dense matter; Equation of state; Stars-neutron}

\section{Introduction}
The data on the glitch
healing parameter, $Q$, provide the best tool for testing the starquake
(Ruderman 1972; Alpar et al 1996) and Vortex unpinning (Alpar et al 1993) models of glitch 
generation in pulsars. Both of these mechanisms of glitch generation, in fact, consider NSs, 
in general, a two component structure: a superfluid interior core surrounded by a rigid crust
 (in the present study we shall use the term `envelope' which includes the solid crust and 
 other interior portion of the star right up to the superfluid core). In the starquake model,
  $Q$ is defined as the fractional moment of inertia, i.e. the ratio of the moment of
inertia of the superfluid core, $I_{\rm core}$, to the moment of inertia of the entire 
configuration, $I_{\rm total}$, as (Pines et al 1974)
\begin{equation}
Q = \frac{I_{\rm core}}{I_{\rm total}}.
\end{equation}
Recently, Crawford \& Demia\'{n}ski (2003) have collected the all measured values of the 
glitch healing parameter $Q$ for Crab and Vela pulsars available in the literature and found 
that for 21 measured values
of $Q$ for Crab glitches, a weighted mean of the values yields $Q = 0.72 \pm 0.05$, and the 
range of $Q 
\geq 0.7$ encompasses the observed distribution for the Crab pulsar. In order to test the 
starquake model for the
Crab pulsar, they have computed $Q$ (as given by Eq.(1)) values for seven representative 
EOSs of dense nuclear matter, covering a range of neutron star masses. 
Their study shows that the much larger values of $Q(\geq 0.7)$ for the Crab pulsar is 
fulfilled by all the six EOSs (out of seven considered in the study) corresponding to a 
`realistic' neutron star mass range 
$1.4\pm 0.2M_\odot$. By contrast, a weighted mean value of the 11 measurements for Vela yields
 a much
smaller value of $Q(= 0.12 \pm 0.07)$ and the all estimates for Vela agree with the likely 
range of $Q \leq 0.2$.
Thus, their results are found to be consistent with the starquake model predictions for the 
Crab pulsar. They 
have also concluded that the much smaller values of $Q \leq 0.2$ for the Vela pulsar are 
inconsistent with the
 starquake model predictions, since the implied Vela mass based upon their models corresponds
  to a value $ \leq 0.5M_\odot$ for $Q \leq 0.2$, which is too low as compared to the 
  `realistic' NS mass range.
  
Thus, in the literature, the starquake is considered as a viable mechanism for glitch
  generation in the Crab and the vortex unpinning, the another mechanism, is considered suitable for the Vela 
  pulsar, since it can avoid some other problems associated with 
 the starquake 
 explanation of the Vela glitches (see, e.g. Crawford \& Demia\'{n}ski (2003); and references
  therein). However, it seems surprising that if the internal structure of NSs are described by 
the same two component conventional  models (as mentioned above), different kinds of glitch 
mechanisms are required for the explanation 
of a glitch! Furthermore, it also follows from the above discussion that the main reason for not considering
the starquake, the feasible mechanism for glitch generation in the Vela, lies in the fact that there 
exists none of the sequence of NS models in the literature which could  explain simultaneously, on the basis of 
starquake model, both the extreme limiting
 cases of glitch healing parameter, $Q$, 
corresponding to the Vela ($Q \leq 0.2$) and the Crab ($Q \geq 0.7$) pulsars in the range 
$0 \leq Q \leq 1.0$  for the `realistic' NS mass values
for both the pulsars. The present study, therefore, deals with the construction 
of such models \footnote{however, the other problems associated with the starquake explanation of the
Vela glitches (see, e.g. Crawford \& Demia\'{n}ski (2003); and references
  therein)
 are not considered in the present paper. The future study in this regard may provide
  some explanation, provided the correlation
between various parameters of the Crab and the Vela pulsar, obtained in the present study,
 can be utilized.}

 We assume that {\em all} the NSs belong to the same family of NS sequence which terminates 
 at the {\em maximum} value of mass. Certainly, this {\em maxima} should correspond to an 
 {\em upper bound} on NS masses. In order to construct such a sequence, we have to set the 
 extreme causal EOS (in geometrized units), $dP/dE = 1$ (where $P$ is the pressure and 
$E$ the energy-density) to describe the core. Firstly, because various observational studies 
like - the gamma-ray burst data, X-ray burst data
and the glitch data etc., and their explanation
(see, e.g. Lindblom 1984; Cottam et al 2002; Datta \& Alpar 1993) favour the stiffest EOSs. 
The latest estimate of the moment of inertia for the Crab pulsar (based upon the
 `newest' observational data on the Crab nebula mass) rules out most of the existing EOSs of 
 the dense nuclear matter, leaving only the stiffest ones (Bejger \& Haensel 2002; Haensel 
 et al 2006). Secondly, because of the fact that the `real' EOS of the dense nuclear matter 
 beyond the density range $\sim 10^{14}\,{\rm g\, cm}^{-3}$
 are largely unknown due to the lack of knowledge of nuclear interactions (see, e.g. Dolan 
 1992; and references therein; Haensel et al 2006), and the various
 EOSs available in the literature (see, e.g. Arnett \& Bowers 1977) for NS matter represent 
 only an extrapolation of the results far beyond this density
 range. Though, the status of the `real' EOS for NS matter is not certain, one could impose 
 some well-known physical principle, independent of the EOS, such as the `causality condition'
  ($dP/dE = 1$) throughout the core of the star beyond a fiduciary density, $E_b$, at the 
  core-envelope boundary to ascertain a definite upper bound on NS masses (see, e.g.,
 Rhoades \& Ruffini 1974; Hartle 1978; Lindblom 1984; Friedman \& Ipser 1987; Kalogera 
 \& Baym 1996). In this connection this is also to be pointed out here that the maximum mass 
 for {\em any} EOS describing the core, beyond the density $E_b$,  with a subluminal sound 
 velocity turns out to be less than that of the upper bound obtained by using the extreme 
 causal EOS (see, e.g., Haensel et al 2006). 

The envelope of our models (below the density $E_b$ at the core-envelope boundary) may be 
characterized by the well-known EOS of classical 
polytrope ${\rm d}$ln$P/{\rm d}$ln$\rho = \Gamma_1$ (where $\rho$ denotes the density of the 
rest-mass and $\Gamma_1$ is a 
constant known as the adiabatic index) for different values of the constant $\Gamma_1 = (4/3), (5/3)$ 
and 2 respectively. The reason for considering the polytropic EOS for the entire envelope lies in the 
fact that with this EOS, our models yield an upper bound on NS masses {\em independent} of the value 
of $\Gamma_1$, and this upper bound (for a fiduciary choice of $E_b$) is found fully consistent with those of the values cited in the literature (Kalogera \& Baym 1996; Friedman \& Ipser 1987). Thus, the choice of the said polytropic EOS for the entire envelope may be regarded entirely equivalent to the choice of the various EOSs like WFF (Wiringa, Fiks \& Fabrocini 1988), FPS (Lorenz, Ravenhall \& Pethick, 1993), 
NV (Negele \& Vautherin 1973), or BPS (Baym, Pethick \& Sutherland 1971) in an appropriate sequence 
below the density range $E_b$, adopted by various authors in the conventional models of NSs 
(see, e.g., Kalogera \& Baym 1996; Friedman \& Ipser 1987), so that the constant $\Gamma_1$ appearing 
in the polytropic EOS may be looked upon as an `average' $\Gamma_1$ for the density range below $E_b$, specified by the sequence of various EOSs in the conventional models of NSs. The choice of the constant $\Gamma_1 = 4/3, 5/3$ and 2 thus become obvious, since this choice can cover almost the entire range of density discussed in the literature for NS matter which is also applicable for the envelope region - the polytropic EOS with $\Gamma_1 = 4/3$ represents the EOS of extreme relativistic degenerate 
 electrons and non-relativistic nuclei (Chandrasekhar 1935), $\Gamma_1 = (5/3)$ represents the well-known EOS of non-relativistic degenerate `neutron gas' (Oppenheimer \& Volkoff 1939), and 
 $\Gamma_1 = 2$ represents the case of extreme relativistic baryons interacting through a vector meson 
 field (Zeldovich 1962) (The value of $\Gamma_1  > 2$ is also  possible 
for some EOS describing the NS matter, e.g.,  Malone,  Johnson \&  Bethe  1975;  Clark, 
Heintzmann \& Grewing 1971, however, the  results  obtained  in 
this paper remain unaffected for the choice of $\Gamma_1  > 2$), and the outcome of this study 
(in terms of explaining the glitch healing parameter for various pulsars and predicting the upper 
bound on the compactness of NSs (since the upper bound on mass is independent of the value of 
$\Gamma_1$)) would finally decide, among the chosen values, the `appropriate' value of $\Gamma_1$ 
for the NS envelope. The validity of assuming the 
 extreme causal EOS in the core and a polytropic EOS in the envelope of the present models, in view of
  the various modern EOS of dense nuclear
 matter, is also discussed in the last section of the present paper.

We have noted that in all conservative models of NSs, the 
choice of the core-envelope boundary, $r_b$ (corresponding to a density denoted by $E_b$), is somewhat 
{\em arbitrary} in the sense that there are no criteria available for the choice of a particular 
matching density, $E_b$, below which the EOS of the NS matter is assumed to be known and unique. One 
can freely choose somewhat lower values of $E_b$ (which will increase the core size) to obtain higher values of $Q$ 
(see, e.g. Shapiro \& Teukolsky 1983; Datta \& Alpar 1993). To avoid such a procedure, we choose the core-envelope boundary of our models on the basis of the `compatibility criterion' which asserts that for an assigned value of the ratio ($\sigma$) of central pressure, $P_0$, to central energy-density, $E_0$, the compactness parameter $u(\equiv M/R$; total mass to radius ratio in geometrized units) of any {\em regular} configuration should not exceed the 
compactness parameter $u_h$ of the homogeneous density sphere, in order to assure the compatibility 
with the hydrostatic equilibrium (Negi \& Durgapal 2001; Negi 2004a). This criterion is capable of constraining 
the core-envelope boundary
 of any physically realistic NS model. A combination of this criterion with those of the observational
  data on the glitch healing parameter and the recently estimated minimum value of the moment of inertia
  for the Crab pulsar (based on the newly estimated `central value' of the Crab nebula mass $M 
  {\rm (nebula)} \simeq 4.6 M_\odot$ in the time-dependent acceleration model), $I_{\rm Crab,45} = 1.93$; where 
  $I_{\rm Crab,45} = I_{\rm Crab}/10^{45}$ g\,cm$^2$ (Bejger \& Haensel 2003)
  can provide the desired NS models discussed above, since both the theory (criterion) and the
  observations (stated above) are being used to construct the NS models.


\section{Methodology}
Since we are using geometrized units (i.e. $G = c = 1$, where $G$ represent the universal constant of gravitation and $c$
is the speed of light in vacuum), the metric for spherically symmetric and static configurations can be written
in the following form
\begin{equation}
ds^2 =  e^{\nu} dt^2 - e^{\lambda} dr^2 - r^2 d\theta^2 - r^2 {\rm sin}^2 \theta d\phi^2 , 
\end{equation}
where $\nu$ and $\lambda$ are functions of $r$ alone. The  Oppenheimer-Volkoff 
(O-V) equations (Oppenheimer \& Volkoff 1939), resulting  from the Einstein field  equations for
systems with isotropic pressure $P$  and  energy-density  $E$  can  be 
written as
\begin{eqnarray}
P' & = & - (P + E)[4 \pi P r^3 + m]/r(r - 2m) \\                       
\nu'/2 & = & - P'/(P + E)  \\                                            
m'(r) & = & 4\pi E r^2 \,;
\end{eqnarray}
where  the prime denotes the radial derivative and $ m(r) $ represents the mass contained within the  radius  $r$

\[m(r)  =  \int_{0}^{r} 4\pi Er^2 dr. \]

The coupled Eqs.(3 - 5) are  solved for the model (supplemented by the boundary conditions:  $P  = E = 0$ , 
$m(r = R)  =  M$, $e^{\nu} = e^{-\lambda} = (1 - 2M/R) = 
(1 - 2u)$ at $r  =  R$)  by considering the EOS, $dP/dE = 1$, in the core and choosing 
various values 
of $\Gamma_1$ in the polytropic envelope for various assigned values of $\sigma$ such that 
for each 
value of $\sigma$, the compactness ratio of the whole configuration always turns out to be 
less than or equal to the compactness ratio of the corresponding sphere (with the same 
$\sigma$) of the
homogeneous density distribution and should not exceed the {\em exact} absolute upper bound 
on compactness ratio of NSs compatible with causality
and pulsational stability (see, e.g. Negi 2004b). The fulfillment of both of the conditions stated in the last sentence is usually called the `appropriate'
fulfillment of `compatibility criterion'. We find that this condition is uniquely 
fulfilled by all the models 
corresponding to an envelope with $\Gamma_1 = (4/3), (5/3)$ and 2 respectively,
if the {\em minimum} value
 of the ratio of pressure to energy-density, $P_b/E_b$, at the core-envelope boundary reaches 
 about $4.694 \times 10^{-2}$. The results of this study are presented in Tables 1 - 2 and Fig.1 for the value 
 of matching density, $E_b = 9.584 \times 10^{14}\,{\rm g\, cm}^{-3}$, i.e. about 3.55 times 
 the nuclear saturation density (note that the particular
choice of $E_b$ used here turns out to be a consequence of the constraints (i) and (ii) 
mentioned in the abstract of the present study, and therefore, does not represent
a fiduciary quantity as discussed in the next section). It is seen that the models 
become pulsationally stable up to the maximum value of mass $M_{\rm max} \simeq 2.22 M_\odot$ and radius, 
$R \simeq 
9.64 - 10.81 {\rm \,km}$. The minimum radius results for the model with a $\Gamma_1 = 2$ 
envelope thus 
maximizes the compactness ratio for the stable configuration, $u \simeq 0.34$, as shown in 
Tables 1 - 2. 
This behaviour indicates that among various values of $\Gamma_1$ chosen in the envelope for 
the density range right from 3.55 times the nuclear saturation density up to zero at the 
surface, the  `average' value of $\Gamma_1$ is appropriately described by a polytropic index 
$n=1$. This fact will also follow from some special features of the model with $\Gamma_1 = 2$ 
envelope discussed in the next section.
The binding energy per unit mass $\alpha [\equiv (M_r - M)/M$; where $M_r$ is the rest-mass (see, e.g. Shapiro \& Teukolsky 1983)] 
also approaches a maximum for about 0.3201 for the maximum value of mass up to which the configurations remain pulsationally stable. However, for the $\Gamma_1 = (4/3)$ envelope model the binding energy reaches a maximum beyond the maximum value of mass.

\begin{table}
      \tbl{ Columns 3 - 5 represent the compactness ratios $u (\equiv M/R)$ of entire configurations for  different assigned values  of $ \sigma \equiv (P_0/E_0) $ (shown in column 1) for the core-envelope models discussed in the text and presented in Tables 2 - 3 and Figs 1 - 3. The core of these models correspond to the stiffest EOS and  the envelope is defined by the polytropic EOS with constant $\Gamma_1 = (4/3), (5/3)$ and 2. The superscripts a, b, and c which appear among the three different values of compactness ratios represent the models corresponding to an envelope with $\Gamma_1 = 4/3, 5/3$, and 2 respectively. Column 2 represents the compactness ratios of homogeneous  
density distribution (indicated by $u_h$) for the same values of $ \sigma$ that correspond to the core-envelope models considered in the present study. The slanted values  correspond  to 
the limiting case up to which the configurations remain pulsationally stable as shown in the $M - R$ diagram (Fig.1) and Table 2 also.}
{\begin{tabular}{@{}ccccc@{}} \toprule
${(P_0 / E_0)}$ & $u_h$  & $u^a$  & $u^b$ & $u^c$ \\ 
\colrule
0.06927 & 0.10813 & 0.04176 & 0.08454 & 0.09544 \\
0.08411 & 0.12531 & 0.04959 & 0.09679 & 0.10925 \\
0.10786 & 0.14970 & 0.06679 & 0.11761 & 0.13131 \\
0.12029 & 0.16116 & 0.07693 & 0.12843 & 0.14236 \\
0.13005 & 0.16960 & 0.08509 & 0.13671 & 0.15077 \\
0.13357 & 0.17253 & 0.08804 & 0.13966 & 0.15370 \\
0.14549 & 0.18205 & 0.09798 & 0.14940 & 0.16340 \\
0.15354 & 0.18814 & 0.10461 & 0.15572 & 0.16969 \\
0.20061 & 0.21911 & 0.14077 & 0.18919 & 0.20255 \\
0.23988 & 0.24008 & 0.16698 & 0.21268 & 0.22536 \\
0.27901 & 0.25763 & 0.18963 & 0.23274 & 0.24469 \\
0.34816 & 0.28259 & 0.22255 & 0.26156 & 0.27253 \\
0.44840 & 0.30929 & 0.25816 & 0.29256 & 0.30222 \\
0.48306 & 0.31666 & 0.26795 & 0.30094 & 0.31024 \\
0.51773 & 0.32332 & 0.27667 & 0.30848 & 0.31746 \\
0.63150 & 0.34115 & 0.29895 & 0.32778 & 0.33589 \\
0.66866 & 0.34593 & {\sl 0.30438} & {\sl 0.33242} & {\sl 0.34040} \\
0.68296 & 0.34766 & 0.30623 & 0.33404 & 0.34194 \\
0.69900 & 0.34952 & 0.30820 & 0.33573 & 0.34352 \\
0.72302 & 0.35220 & 0.31076 & 0.33793 & 0.34567 \\
0.74010 & 0.35401 & 0.31235 & 0.33934 & 0.34700 \\
0.77503 & 0.35751 & 0.31501 & 0.34164 & 0.34914 \\
0.80605 & 0.36041 & 0.31656 & 0.34294 & 0.35044 \\
0.83972 & 0.36336 & 0.31711 & 0.34347 & 0.35091 \\
\botrule
\end{tabular} \label{ta1}}
\end{table}

\begin{table*}
      \tbl{Mass ($M$), size ($R$) and surface redshift ($z_R$) for different values of the ratio of central pressure to central energy-density, $(P_0/E_0)$, for the core-envelope models discussed in the text. Various parameters are  obtained  by  assigning the {\it calculated} value of $E_b = 9.584 \times 10^{14}$ g\, cm$^{-3}$ and the {\it minimum} value of the ratio of pressure to energy-density, $(P_b/E_b) \simeq 4.694 \times 10^{-2}$, at the core-envelope boundary. This {\it minimum} value of $(P_b/E_b)$ is obtained in such a manner that together with the `compatibility criterion', $u \leq
u_h$  (where $u_h$  represents the  compactness ratio of homogeneous density sphere for the corresponding value of $P_0/E_0$ as shown in Table 1), the absolute upper bound on compactness ratio compatible with causality and dynamical stability, $0.3406 \leq u$ (Negi 2004b), also turn out to be satisfied for {\it all} the models corresponding to an envelope with $\Gamma_1 = 4/3, 5/3$, and 2 respectively (that is, the compatibility criterion is `appropriately' satisfied). The superscripts a, b, and c which appear among various parameters represent the models corresponding to an envelope with $\Gamma_1 = 4/3, 5/3$, and 2 respectively. The 
slanted values correspond to the limiting case upto which the configurations 
remain pulsationally stable.}
{\begin{tabular}{@{}cccccccccc@{}} \toprule
${(P_0 / E_0)}$  & $(M^a/M_{\odot})$    & $R^a({\rm km})$ & $z_R^a$ & $(M^b/M_{\odot})$  & $R^b({\rm km})$   & $z_R^b$ & $(M^c/M_{\odot})$   & $R^c({\rm km})$ & $z_R^c$ \\ 
\colrule
0.06927 & 0.89789 & 31.75378 & 0.04458 & 0.66883 & 11.68584 & 0.09703 & 0.59951 & 9.27742 & 0.11172 \\
0.08411 & 0.90891 & 27.07333 & 0.05361 & 0.73848 & 11.26932 & 0.11357 & 0.68456 & 9.25509 & 0.13119 \\
0.10786 & 0.97639 & 21.59094 & 0.07433 & 0.86614 & 10.87749 & 0.14349 & 0.82806 & 9.31398 & 0.16454 \\
0.12029 & 1.02595 & 19.69704 & 0.08713 & 0.93555 & 10.75926 & 0.16002 & 0.90301 & 9.36878 & 0.18239 \\
0.13005 & 1.06836 & 18.54447 & 0.09776 & 0.98995 & 10.69557 & 0.17316 & 0.96091 & 9.41313 & 0.19655 \\
0.13357 & 1.08412 & 18.18706 & 0.10169 & 1.00944 & 10.67591 & 0.17795 & 0.98151 & 9.43181 & 0.20160 \\
0.14549 & 1.13857 & 17.16337 & 0.11522 & 1.07461 & 10.62370 & 0.19421 & 1.04998 & 9.49115 & 0.21878 \\
0.15354 & 1.17576 & 16.60133 & 0.12453 & 1.11771 & 10.60136 & 0.20512 & 1.09496 & 9.53069 & 0.23034 \\
0.20061 & 1.38720 & 14.55530 & 0.17977 & 1.35108 & 10.54794 & 0.26834 & 1.33583 & 9.74097 & 0.29651 \\
0.23988 & 1.54530 & 13.66884 & 0.22532 & 1.51885 & 10.54804 & 0.31917 & 1.50724 & 9.87834 & 0.34929 \\
0.27901 & 1.68274 & 13.10652 & 0.26925 & 1.66237 & 10.54976 & 0.36778 & 1.65320 & 9.97906 & 0.39943 \\
0.34816 & 1.87902 & 12.47035 & 0.34244 & 1.86515 & 10.53233 & 0.44809 & 1.85870 & 10.07341 & 0.48259 \\
0.44840 & 2.07419 & 11.86706 & 0.43787 & 2.06530 & 10.42657 & 0.55254 & 2.06105 & 10.07287 & 0.58997 \\
0.48306 & 2.12116 & 11.69220 & 0.46790 & 2.11337 & 10.37226 & 0.58488 & 2.10963 & 10.04365 & 0.62323 \\
0.51773 & 2.15898 & 11.52577 & 0.49627 & 2.15211 & 10.30421 & 0.61577 & 2.14879 & 9.99748  & 0.65501 \\
0.63150 & 2.22484 & 10.99220 & 0.57700 & 2.22007 & 10.00366 & 0.70392 & 2.21773 & 9.75186 & 0.74551 \\
0.66866 & {\sl 2.22812} & {\sl 10.81174} & {\sl 0.59876}  & {\sl 2.22383} &  {\sl 9.88101} & {\sl 0.72730}  &  {\sl 2.22173} & {\sl 9.64004} & {\sl 0.76999} \\
0.68296 & 2.22694 & 10.74105 & 0.60634  & 2.22282 & 9.82845  & 0.73574 & 2.22080 & 9.59274 & 0.77857 \\
0.69900 & 2.22411 & 10.65856 & 0.61460 &  2.22016 & 9.76720  & 0.74466 & 2.21822 & 9.53752 & 0.78753 \\
0.72302 & 2.21647 & 10.53475  & 0.62545 & 2.21277 & 9.67152  & 0.75642 & 2.21095 & 9.44716 & 0.79993 \\
0.74010 & 2.20849 & 10.44318  & 0.63235 & 2.20495 & 9.59713  & 0.76415 & 2.20321 & 9.37783 & 0.80777 \\
0.77503 & 2.18547 & 10.24715  & 0.64403 & 2.18222 & 9.43445  & 0.77687 & 2.18062 & 9.22490 & 0.82053 \\
0.80605 & 2.15636 & 10.06098  & 0.65099 & 2.15335 & 9.27424  & 0.78422 & 2.15186 & 9.06952 & 0.82840 \\
0.83972 & 2.11385 & 9.84559  & 0.65346 &  2.11104 & 9.07808  & 0.78723 & 2.10966 & 8.87967  & 0.83131 \\
\botrule
\end{tabular} \label{ta1}}
\end{table*}

\section {An application of the models to the Crab and Vela pulsars}

For slowly rotating configurations like the Crab and the Vela pulsars (rotation velocity about 188 and 70 rad sec$^{-1}$ respectively) the moment 
of inertia may be calculated in the first order approximation that appears in the form of the Lense-Thirring 
frame dragging-effect. For Crab and Vela pulsars, the first-order effects turn out to be about 1 - 2\%  (other effects
like mass shift and deformation from spherical symmetry due to rotation represent second-order effects which are significant for
the case of millisecond pulsars. For Crab and Vela like pulsars, the second-order effects turn out to be about $10^{-4}$ or
even lower; see, e.g. Arnett \& Bowers 1977; Crawford \& Demia\'{n}ski 2003. Therefore, these effects can be safely ignored when studying the 
macroscopic parameters of the slowly
rotating pulsars as carried out in the present paper). These effects are reproduced by an empirical formula 
that is based on the numerical results obtained for thirty theoretical EOSs of dense
nuclear matter. For NSs, the formula yields (Bejger \& Haensel 2002)
\begin{equation}
I \simeq \frac{2}{9}(1 + 5x)MR^2, \,\,\,\, x > 0.1 
\end{equation}

where $x$ is the compactness ratio measured in units of $(M_\odot({\rm km})/{\rm km})$, i.e.
\begin{equation}
x = \frac{M/R}{M_\odot/{\rm km}} = \frac{u}{1.477}
\end{equation}
only static (non-rotating) parameters of the spherical configuration
appear in the formula.

Equation (6) is used, together with coupled Eqs.(3 - 5), to calculate the fractional moment of
 inertia
given by Eq.(1) and the moment of inertia of the entire configuration for an assigned value 
of the matching density, $E_b$. The results of the calculations are presented in Table 3 and Figs. 2 - 3.
It is seen from Table 3 that among all models corresponding to different 
values of $\Gamma_1$ in the envelope, the lower limits of both the observational constraints 
for the Crab pulsar, $Q \geq 0.7$ and $I_{\rm Crab,45} \equiv I_{\rm total,45} \geq 1.93$, 
are appropriately satisfied by the $\Gamma_1 = 2$ envelope model. Since this particular model
 yields a unique value of matching density, $E_b = 9.584 \times 10^{14}$ g\, cm$^{-3}$, for 
 the last two constraints in such a manner that the constraint $I_{\rm total,45} \geq 1.93$ 
 is appropriately fulfilled by all stable NS models corresponding to the $\Gamma_1 = (5/3)$ 
 and 2 envelopes for another constraint $Q \geq 0.7$ as shown in Table 3 and Figs.2 - 3 respectively 
 (the model with a $\Gamma_1 = (4/3)$ envelope is ruled out since it always yields stable 
 models with $Q < 0.7$). Another special feature of the $\Gamma_1 = 2$ envelope model lies 
 in the fact that it yields somewhat wider range of glitch healing parameter, $0 < Q \leq 0.78
 $, for the stable models as compared to the $\Gamma_1 = (5/3)$ envelope model which yields 
 this parameter in the range $0 < Q \leq 0.75$ for stable models as shown in Fig.2 and Tables 2 - 3. 
 Furthermore, the $\Gamma_1 = 2$ envelope model yields somewhat higher value of maximum 
 surface redshift, $z_R \simeq 0.2016$, for the Vela pulsar ($Q \simeq 0.2$) and somewhat 
 lower value of minimum surface redshift, $z_R \simeq 0.6232$, for the Crab pulsar ($Q \simeq
  0.7$) simultaneously, as compared to the $\Gamma_1 = (5/3)$ envelope model as shown in Tables 2 - 3. 
Thus provides somewhat wider range of lower and upper bounds on the energy of a gravitationally redshifted
   radiation from the surface of the Vela and the Crab pulsars respectively. And above all, 
   the compactness maximizes, $u \simeq 0.34$, for the $\Gamma_1 = 2$ envelope model in the 
   stable sequence of NS models as shown in Tables 1 - 2 and Fig.1. This special feature of the model is also 
   discussed in section 2.

\begin{table*}
      \tbl{Fractional moment of inertia, $Q(\equiv I_{\rm core}/I_{\rm total)}$, and the moment of inertia of the entire configuration, $I_{\rm total,45}$, for different values of the ratio of pressure to energy-density, $(P_0/E_0)$, at the centre for the  models presented in Table 2. As stated in Table 2, these parameters are obtained for the {\it minimum} value of the ratio of pressure to energy-density, $(P_b/E_b) \simeq 4.694 \times 10^{-2}$ and for the calculated value of $E_b  = 9.584 \times 10^{14}$ g\, cm$^{-3}$ at the core-envelope boundary. The superscripts a, b, and c which appear among various parameters represent the models corresponding to an envelope with $\Gamma_1 = 4/3, 5/3$, and 2 respectively. The 
slanted values correspond to the limiting case upto which the configurations 
remain pulsationally stable. In other words, we can state that since both of the observational constraints ($Q \simeq 0.7$ 
 and $I_{\rm Crab,45} \simeq 1.93$ imposed by the Crab pulsar) yields a boundary density, $E_b  = 9.584 \times 10^{14}$ g\, cm$^{-3}$, for the $\Gamma_1 = 2$ envelope model at the core-envelope boundary, $P_b/E_b \simeq 4.694 \times 10^{-2}$, which, in fact, is assured on the basis of `appropriate' satisfaction of the `compatibility criterion' for {\it all} the sequences corresponding to  $\Gamma_1 = 4/3, 5/3$, and 2 envelope models. It follows, therefore, that  $E_b  = 9.584 \times 10^{14}$ g\, cm$^{-3}$ represents a unique value of boundary density applicable for all the sequences corresponding to  $\Gamma_1 = 4/3, 5/3$, and 2 envelope models.}
{\begin{tabular}{@{}ccccccc@{}} \toprule
${(P_0 / E_0)}$ &  $Q^a$ & $I^a_{\rm total,45}$ & $Q^b$ &  $I^b_{\rm total,45}$ & $Q^c$ & $I^c_{\rm total,45}$ \\ 
\colrule
0.06927 & 0.00118 & 4.57097 & 0.01036 & 0.51964 & 0.01783 & 0.30200 \\
0.08411 & 0.00522 & 3.44158 & 0.03265 & 0.55078 & 0.05061 & 0.35531 \\
0.10786 & 0.02242 & 2.46867 & 0.08733 & 0.63381 & 0.12059 & 0.45901 \\
0.12029 & 0.03739 & 2.21928 & 0.12071 & 0.68735 & 0.15969 & 0.51958  \\
0.13005 & 0.05168 & 2.09336 & 0.14763 & 0.73277 & 0.19016 & 0.56886 \\
0.13357 & 0.05729 & 2.05901 & 0.15739 & 0.74953 & 0.20090 & 0.58720  \\
0.14549 & 0.07777 & 1.97575 & 0.19020 & 0.80784 & 0.23645 & 0.64982 \\
0.15354 & 0.09261 & 1.94099 & 0.21182 & 0.84859 & 0.25950 & 0.69269  \\
0.20061 & 0.18555 & 1.91950 & 0.32652 & 1.09079 & 0.37684 & 0.94514  \\
0.23988 & 0.26028 & 1.99907 & 0.40470 & 1.28571 & 0.45366 & 1.14695  \\
0.27901 & 0.32671 & 2.09949 & 0.46877 & 1.46324 & 0.51517 & 1.33145  \\
0.34816 & 0.42326 & 2.26638 & 0.55590 & 1.72560 & 0.59803 & 1.60402  \\
0.44840 & 0.52531 & 2.42132 & 0.64342 & 1.97684 & 0.67967 & 1.87142 \\
0.48306 & 0.55283 & 2.44624 & 0.66607 & 2.03036 & 0.70071 & 1.93000 \\
0.51773 & 0.57715 & 2.45692 & 0.68625 & 2.06633 & 0.71944 & 1.97100 \\
0.63150 & 0.63925 & 2.39256 & 0.73770 & 2.07327 & 0.76712 & 1.99374 \\
0.66866 & {\sl 0.65474} & {\sl 2.33926} & {\sl 0.75034} & {\sl 2.04122} &  {\sl 0.77916} & {\sl 1.96573} \\
0.68296 & 0.66008 & 2.31463 & 0.75494 & 2.02387 & 0.78336 & 1.95037 \\
0.69900 & 0.66592 & 2.28380 & 0.75976 & 2.00170 & 0.78778 & 1.93052  \\
0.72302 & 0.67364 & 2.23279 & 0.76625 & 1.96294 & 0.79403 & 1.89425  \\
0.74010 & 0.67868 & 2.19200 & 0.77068 & 1.93034 & 0.79815 & 1.86389  \\
0.77503 & 0.68772 & 2.09761 & 0.77855 & 1.85289 & 0.80543 & 1.79104  \\
0.80605 & 0.69415 & 2.00024 & 0.78427 & 1.77042 & 0.81111 & 1.71182  \\
0.83972 & 0.69887 & 1.87943 & 0.78917 & 1.66437 & 0.81587 & 1.60991 \\
\botrule
\end{tabular} \label{ta1}}
\end{table*}

The above-mentioned features indicate the appropriateness of the $\Gamma_1 = 2$ envelope model
 and thus emerges an important consequence of this study which states that the envelope of 
 `real' NSs may be well approximated by an EOS of classical polytrope with `average' value 
 of polytropic index, $n$, closer to 1.

For the minimum value of $Q \simeq 0.7$, Table 3 yields $I_{\rm total,45} \equiv I_{\rm Crab,45}
 \simeq 1.93$ for the $\Gamma_1 = 2$ envelope model and from Table 2, on the basis of $\Gamma_1 = 2$ envelope model,
we obtain the minimum values of mass, $M \simeq 2.11 M_\odot$, and surface redshift, $z_R \simeq
  0.6232$, for the Crab pulsar. On the other hand, for the maximum value of $Q \simeq 0.2$ 
  which belongs to the Vela pulsar, Table 3 and Fig.3 yield $I_{\rm total,45} \simeq 0.587$ corresponding
   to the $\Gamma_1 = 2$ envelope model  and Table 2 and Fig.2 yield the maximum values of 
   mass, $M \simeq 0.982 M_\odot$, and surface redshift, $z_R \simeq 0.2016$, for the Vela 
   pulsar (note that the upper weighted mean value of $Q \approx 0.19$ for the Vela pulsar, 
    Table 2 and Fig.2(for the $\Gamma_1 = 2$ envelope model) yield the maximum value of mass,
    $M \simeq 0.961 M_\odot$, and surface redshift, $z_R \simeq 0.1966$, for the Vela pulsar. 
    Obviously, these values are much closer to the values corresponding to the maximum value 
    of $Q \approx 0.2$ for the Vela pulsar). However, for the observational constraint of the
     `central' weighted mean values of $Q \approx 0.72$ for the Crab and $Q \approx 0.12$ for
      the Vela pulsar, Table 2 and Fig.2 (for the $\Gamma_1 = 2$ envelope model) yield the 
      slightly increased values of mass, $M \simeq 2.149 M_\odot$, and the surface redshift, 
      $z_R \simeq 0.655$, for the Crab and somewhat decreased values of mass, $M \simeq 0.828
       M_\odot$, and surface redshift, $z_R \simeq 0.1645$, for the Vela pulsar respectively. For 
       these central weighted mean values of $Q$, Tables 2 - 3 and Fig.3 yield for the $\Gamma_1 = 2$ envelope model, 
       the moment of inertia $I_{\rm Crab,45} \simeq 1.971$ for the Crab and $I_{\rm Vela,45} 
       \simeq 0.459$ for the Vela pulsar respectively.
For the lower weighted mean value of $Q \approx 0.05$ corresponding to the Vela pulsar, Tables 2 - 3 and Fig.2 
 yield (for the $\Gamma_1 = 2$ envelope model) the minimum values of mass, 
$M \simeq 0.685 M_\odot$, and surface redshift, $z_R \simeq 0.1312$, whereas Table 3 and Fig.3 yield a 
minimum value of moment of inertia, $I_{\rm Vela,45} \simeq 0.355$, for the Vela pulsar.

\section{Implications of the models for the extraordinary gamma-ray burst GRB790305b}   

Apart from the models of the Crab and Vela pulsars, let us consider two of the main findings related to the 
extraordinary gamma-ray burst of 5 March 1979: (i) it gives the only reliable estimate of the surface redshift,
$z_R = 0.23\pm 0.07$ ( after taking due account of thermal blueshift), associated with the supernova remnant N49
in the Large Magellanic Cloud (see, e.g. Higdon \& Lingenfelder 1990; Douchin \& Haensel 2001; and references therein),
and (ii) the implied peak luminosities of the repeating burst GRB790305b correspond to an energy of $10^{44}$ ergs,
which is possible only when a starquake releases at least $10^{-9}$ of the NS gravitational binding-energy of $10^{53}$ ergs
(Higdon \& Lingenfelder 1990). In order to reproduce both of these findings,
one would require a NS model compatible with starquake model predictions that could also account
for a surface redshift $z_R = 0.23\pm0.07$. Both of these requirements are, in fact, fulfilled by
our models. For the case of the $\Gamma_1 = 2$ envelope model, we get from Tables 2- 3 and Fig. 2 - 3 the central value
of surface redshift $z_R = 0.23$ for a mass $M \simeq 1.095 M_\odot$ with $Q \simeq 0.260$. 
The binding energy corresponding to this case is obtained as $2.618 \times 10^{53}$ ergs which is
capable of releasing $2.618 \times 10^{44}$ ergs of energy required for the latter burst.

\section{Results and conclusions}

This study constructs the stable sequences of NS models terminate at the value of maximum mass, $M_{\rm
max} \simeq 2.22 M_\odot$, independent of the EOSs of the envelope, for the matching density, 
$E_b = 9.584 \times 10^{14}$ g\, cm$^{-3}$, at the core-envelope 
boundary. This value of `matching density' is a consequence of the observational constraints $Q \simeq 0.7$ and $I_{\rm Crab,45} \simeq 1.93$ 
(associated with the Crab pulsar) imposed together on the $\Gamma_1 = 2$ envelope model and in this sense
does not represent a fiduciary quantity. The upper bound of the surface redshift, $z_R \simeq 0.77$ (corresponding
to a $u$ value $\simeq 0.34$), however, belongs to the model with a $\Gamma_1 = 2$ envelope which is consistent
with the absolute upper bound on the surface redshift of NS models compatible with causality and pulsational stability (Negi 2004b). 
This special feature, together with some other remarkable ones, discussed in the present study underline the appropriateness of the $\Gamma_1 = 2$ envelope model. 

Since among the variety of modern EOSs discussed in the 
 literature, the upper bound on NS mass compatible with causality and dynamical stability can reach 
 a value up to $2.2M_\odot$ (in this category, the SLy 
 (Douchin \& Haensel 2001) EOS yields a maximum mass of $2.05M_\odot$, whereas the BGN1 (Balberg \& Gal 1997) 
 and the APR (Akmal et. al. 1998) EOSs yield the maximum masses of $2.18M_\odot$ and $2.21M_\odot$ respectively (see, e.g. Haensel et al 2006)).
 In view of this result, the model-independent maximum mass, $M_{\rm max} \simeq 2.22M_\odot$, obtained 
in this study may be regarded as good as those obtained on the basis of modern nuclear theory.

In addition to this result, the appropriate sequences of stable NS models obtained in this study can explain the glitch 
healing parameter, $Q$, of any glitching pulsar, provided the weighted mean values of $Q$ 
lie in the range $0 < Q \leq 0.78$. This finding also reveals that if the starquake is 
considered to be a viable mechanism for glitch generation in all pulsars, then the envelope 
of `real' NSs may be well approximated by a polytropic EOS corresponding to a polytropic 
index, $n$, closer to 1.

For the value of matching density, $E_b = 9.584 \times 10^{14}$ g\, cm$^{-3}$, the $\Gamma_1 = 2$ envelope
model yields the minimum values of mass $M \simeq 2.11 
M_\odot$ and surface redshift $z_R \simeq 0.6232$ for the Crab ($Q \simeq 0.7$) and the maximum values 
of mass $M \simeq 0.982 M_\odot$ and surface redshift $z_R \simeq 0.2016$ for the Vela pulsar ($Q \simeq
 0.2$). The minimum mass and surface redshift for the Crab pulsar are slightly increased up to the values 
 $M \simeq 2.149 M_\odot$ and $z_R \simeq 0.655$ respectively, if the `central' weighted mean value of 
 $Q \approx 0.72$ and the moment of inertia $I_{\rm Crab,45} >
1.93 $ are also 
 imposed on these models. However, for the `central' weighted mean value of $Q \simeq 0.12$ 
 corresponding to the Vela pulsar, the maximum mass and surface redshift are somewhat decreased to 
 the values $M \simeq 0.828 M_\odot$ and $z_R \simeq 0.1645$ respectively. This value of mass and 
 surface redshift for the Vela pulsar can further decrease up to the values $M \simeq 0.685 M_\odot$ 
 and $z_R \simeq 0.1312$ respectively, if the lower weighted mean value of $Q \simeq 0.05$ for the 
 Vela pulsar is imposed. These results predict the upper and lower bounds on the energy of a 
 gravitationally redshifted electron-positron
annihilation line in the range of about 0.309 - 0.315 MeV from the Crab and in the range of about 
0.425 - 0.439 MeV from the Vela pulsar respectively.

For a comparison, if the observational constraint of the minimum value of $I_{\rm Crab,45} \simeq 3.04$
(the value of moment of inertia for the Crab pulsar obtained earlier by Bejger \& Haensel (2002), on the 
basis of the constant-acceleration model for the Crab nebula) together with $Q \simeq 0.7$ is imposed
on the models studied in the present paper, the $\Gamma_1 = 2$ envelope model yields the value of matching density, 
$E_b = 7.0794 \times 10^{14}$ g\, cm$^{-3}$. This value of $E_b$ yields a model-independent upper bound on NS mass
$M_{\rm max} \simeq 2.59 M_\odot$. This value of maximum mass, however, represents an `average' of the
 maximum NS masses in the range $2.2M_\odot \leq M_{\rm max} \leq 2.9M_\odot$ obtained by Kalogera \& Baym 1996
(and references therein) on the basis of other EOSs for NS matter, fitted to experimental 
nucleon-nucleon scattering data and the properties of light nuclei.
For this lower value of matching density, the $\Gamma_1 = 2$ envelope models yield the minimum value of
 mass $M \simeq 2.455 M_\odot$ for the Crab ($Q \simeq 0.7$) and the maximum value 
of mass $M \simeq 1.142 M_\odot$ for the Vela pulsar ($Q \simeq 0.2$). The minimum mass for the 
Crab pulsar is slightly increased up to the value $M \simeq 2.5 M_\odot$, if the `central' weighted mean value of 
 $Q \approx 0.72$ and the moment of inertia $I_{\rm Crab,45} > 3.04 $ are also 
 imposed on these models. However, corresponding to the `central' weighted mean value of $Q \simeq 0.12$,
  the maximum mass of the Vela pulsar is somewhat decreased to 
 the value $M \simeq 0.964 M_\odot$. This value of mass for the Vela pulsar can further decrease up 
 to the value $M \simeq 0.796 M_\odot$, if the lower weighted mean value of $Q \simeq 0.05$ for the 
 Vela pulsar is imposed.
 
Furthermore, the study can also explain some special features associated with the extraordinary 
gamma-ray burst of 5 March 1979.

\section*{Acknowledgments}

The author acknowledges the  Aryabhatta Research Institute of Observational Sciences (ARIES), Nainital 
for providing library and computer-centre facilities.
\begin{figure}
  \centering
   \includegraphics[height=9.5cm,width=9.5cm]{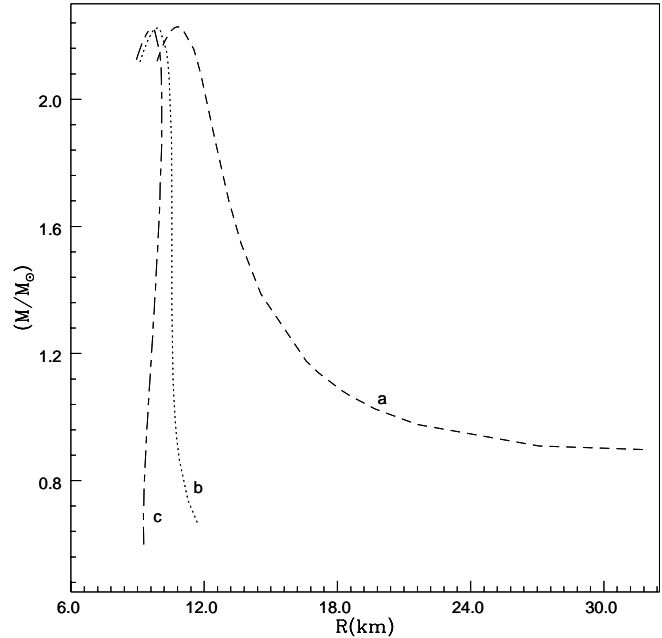}

      \caption{Mass-Radius diagram of the models as discussed in the text and presented in Table 2 for the value of matching density
            $E = E_b = 9.584 \times 10^{14}$ g\, cm$^{-3}$ at the core-envelope
            boundary. The labels a, b and c represent the models for an envelope with $\Gamma_1 =  (4/3), 
            (5/3)$ and 2 respectively. The  minimum  value  of  the  ratio  of 
pressure to energy-density, $(P_b/E_b)$, at the core envelope boundary is obtained as $4.694 \times 10^{-2}$, such that for an assigned value of $ \sigma $, the inequality $u \leq u_h $  is always satisfied for {\em all} the models as shown in Table 1.
              }
        \label{FigVibStab}
  \end{figure}

  \begin{figure}
   \centering
     \includegraphics[height=9.5cm,width=9.5cm]{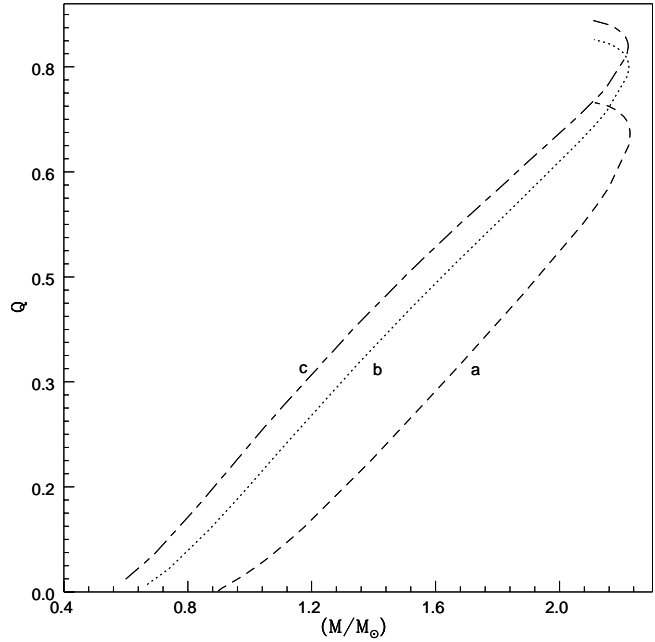}   

      \caption{Fractional moment of inertia $Q (\equiv I_{\rm core}/I_{\rm total})$ vs. total mass $M$ (in units of $M_\odot$)
      for the configurations
      presented in Tables 2 - 3 and Fig.1. The labels a, b and c represent the models for an envelope with $\Gamma_1 =  (4/3), 
            (5/3)$ and 2 respectively.
              }
 
 \label{FigVibStab}             
 
 \end{figure}

\vspace{1.0cm}


   \begin{figure}
   \centering
     \includegraphics[height=9.5cm,width=9.5cm]{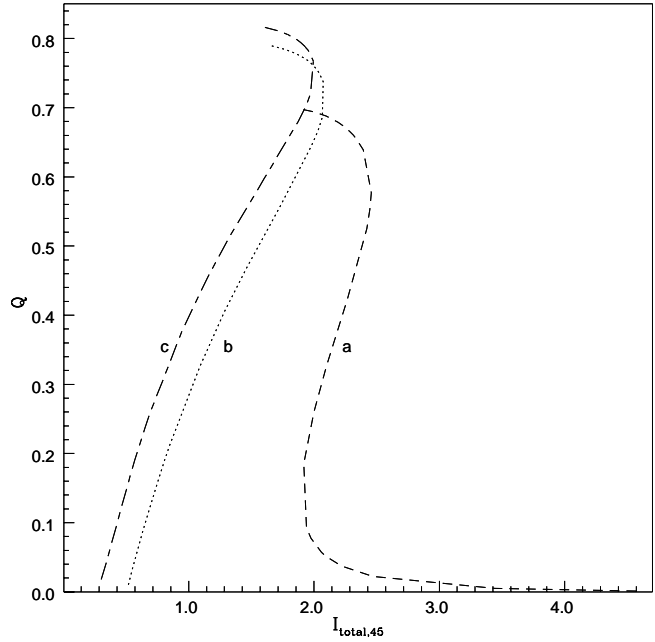}

      \caption{Fractional moment of inertia $Q (\equiv I_{\rm core}/I_{\rm total})$ vs. moment of inertia
      of the entire structure $I_{\rm total,45}$
      for the configurations
      presented in Tables 2 - 3 and Figs.1 - 2. The labels a, b and c represent the models for an envelope with $\Gamma_1 =  (4/3), 
            (5/3)$ and 2 respectively. $I_{\rm total,45}$ is defined as $I_{\rm total}/10^{45} {\rm g\, cm^2}$.
              }
          
         \label{FigVibStab}
   \end{figure}

\end{document}